\newif\ifjpsj
\jpsjfalse

\ifjpsj
 \documentclass[twocolumn,letter]{jpsj3}
\else
 \documentclass[twocolumn,amsmath,amssymb,longbibliography,prb]{revtex4-2}
 \usepackage[dvipdfmx]{graphicx}
 \usepackage[pdftex]{hyperref}
 \hypersetup{
	colorlinks=true
 }
\fi

\usepackage{txfonts}

\usepackage{bm}
\usepackage{color}
\usepackage{braket}
\usepackage{mathdots}
\usepackage[version=3]{mhchem} 
\usepackage{hyperref}

\newif\iftitle
\titletrue

\renewcommand{\(}     {\left(}
\renewcommand{\)}     {\right)}
\renewcommand{\[}     {\left[}
\renewcommand{\]}     {\right]}
\renewcommand{\_}[1]  {_\textrm{#1}}

\ifjpsj
 \newcommand{\onlinecite}[1]     {\citen{#1}}
\fi

\begin{document}

\title{
Intrinsic and Extrinsic Anomalous Hall Effects in Disordered Magnetic Weyl Semimetal
}

\ifjpsj
 \author{Koji Kobayashi$^1$\thanks{k-koji@tohoku.ac.jp} and Kentaro Nomura$^{1,2}$}
 \inst{
 $^1$Institute for Materials Research, Tohoku University, Sendai 980-8577, Japan\\
 $^2$Center for Spintronics Research Network, Tohoku University, Sendai 980-8577, Japan
 }
\else
 \author{Koji Kobayashi$^1$\thanks{k-koji@tohoku.ac.jp}}
 \author{Kentaro Nomura$^{1,2}$}
 \affiliation{$^1$Institute for Materials Research, Tohoku University, Sendai 980-8577, Japan}
 \affiliation{$^2$Center for Spintronics Research Network, Tohoku University, Sendai 980-8577, Japan}
\fi

\newcommand{\abstbody}{
We study the intrinsic and extrinsic Hall effects in disordered magnetic Weyl semimetals numerically.
 We show that in Weyl metals, where the Fermi energy deviates from the Weyl point, 
the Hall and longitudinal conductances exhibit a specific relation, 
which is distinguished from the well-known relation in integer quantum Hall systems.
 Around the Weyl point, the Hall conductance increases with increasing longitudinal conductance.
 This increasing behavior indicates the existence of additional contributions to the Hall conductance from the subbands of Weyl cones besides that from the bulk Berry curvature.
 We also show that the extrinsic anomalous Hall effect due to the spin scatterers (skew scattering) is significantly suppressed in Weyl metals.
}
\ifjpsj
 \abst{\abstbody}
\else
 \begin{abstract}
  \abstbody
 \end{abstract}
\fi

\maketitle


\iftitle
\section{Introduction} \label{sec:intro}
\else
 \textit{Introduction.}
\fi
 The Hall effect arises in the systems under an external magnetic field or those with both magnetic ordering and spin-orbit coupling; they are called the ordinary Hall effect and anomalous Hall effect, respectively \cite{Nagaosa10anomalous}.
 Recently, the Hall effect due to the topological origin, which arises in Weyl semimetals (WSMs)\cite{Burkov14anomalous,Armitage18weyl} or magnetic topological insulators \cite{Yu10quantized,Tokura19magnetic}, renewed the interest in the Hall effect.
 The origin and properties of the Hall effect have been intensively studied not only for their theoretical importance but also for device applications such as magnetic sensors \cite{Heremans93solid,VilanovaVidal11exploring,Satake19fe,Fujiwara19doping}.
 It is known that depending on the origin of the Hall effect, the Hall conductance and longitudinal conductance show specific relations.
 For example, the impurity-driven (skew scattering) extrinsic Hall conductance in ordinary metals is proportional to the longitudinal conductance \cite{Nagaosa10anomalous}, 
while the Hall conductance in disordered quantum Hall insulators (QHIs) shows a dome-shaped relation to the longitudinal conductance \cite{Aoki87quantized,Huckestein95scaling}.
 Therefore, it is a fundamental problem to reveal the relation between the Hall and longitudinal conductances for understanding the Hall effect.

 However,
the relation of the conductances in WSMs is still unclear.
 One of the reasons is that the WSMs show non-monotonic Hall conductance as a function of the Fermi energy \cite{Burkov14anomalous,Burkov15chiral}.
 Another reason is the technical difficulty.
 To evaluate the longitudinal conductance,
we need a large-scale calculation with a sufficiently strong disorder,
which is numerically challenging.
 We overcome this difficulty by investigating thin films of WSM and their size dependence instead of starting from the three-dimensional limit.

 In this 
\iftitle
 paper,
\else
 Letter, 
\fi
we study the relation between the Hall and longitudinal conductances in disordered WSMs.
 The conductances are calculated by the Landauer-B\"{u}ttiker formula in lattice models%
\iftitle
 introduced in Sec.~\ref{sec:model}%
\fi.
 We first demonstrate the QHI-like conductance relation in disordered WSM thin films at the Weyl point%
\iftitle
 in Sec.~\ref{sec:QuantizedRegime}%
\fi.
 Then 
we show the conductances as functions of energy and their size dependence%
\iftitle
 in Sec.~\ref{sec:EnergyDependence}%
\fi.
\iftitle
 In Sec.~\ref{sec:extrinsic},
\else
 In addition, 
\fi
we confirm the suppression of extrinsic (i.e., impurity induced) contribution to the Hall conductance in WSMs.
\iftitle
 The conclusion is given in Sec.~\ref{sec:conclusion}.
\fi

\begin{figure}[!htbp]
 \centering
  \includegraphics[width=\linewidth]{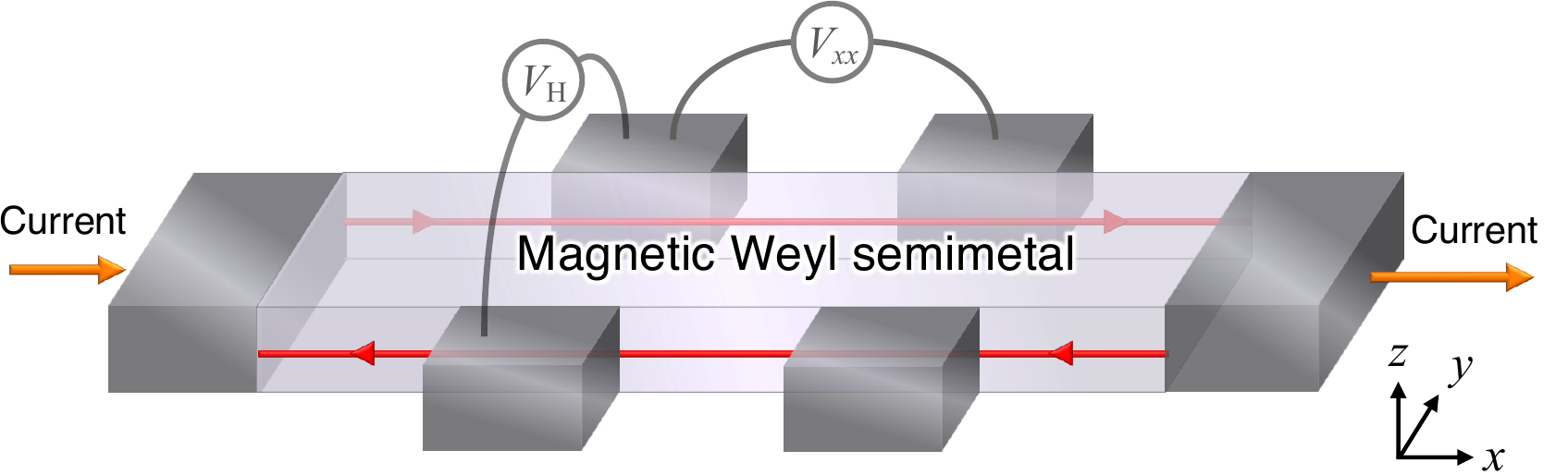}
 \vspace{-3mm}
\caption{\ifjpsj(Color online) \fi
  Schematic figure of the system geometry.
  The system size is $2L \times L \times N$.
  Voltage probes of the width $2L/5$ are attached on the surfaces where chiral surface states emerge.
  The separations between the probes are $2L/5$.
  In the $z$ direction, the periodic boundary conditions are imposed for simplicity.
}
\label{fig:geometry}
\end{figure}


\iftitle
\section{Weyl semimetal model} \label{sec:model}
\else
 \textit{Weyl semimetal model.}
\fi
 We employ a simple model for magnetic WSMs on the cubic lattice,
\cite{Yang11quantum,Imura11spin,Chen15disorder,Shapourian16phase,Liu16effect,Yoshimura16comparative}
\begin{align}
 H\_{W} &=  \sum_{\bf r}
         \[ {it \over 2} 
           \(
              \ket{{\bf r}+{\bf e}_x}
               \sigma_x
              \bra{\bf r}
            + \ket{{\bf r}+{\bf e}_y}
               \sigma_y
              \bra{\bf r}
           \)
           + \textrm{H.c.}
         \] \nonumber \\
   &+  \sum_{\bf r}
        \sum_{\mu=x,y,z}
         \[ \ket{{\bf r}+{\bf e}_\mu}
           \(
              -{m_2 \over 2} \sigma_z
           \)
           \bra{\bf r}  + \textrm{H.c.}
         \]   \nonumber \\
   & + \sum_{\bf r} \ket{\bf r}
        \[m_0 \sigma_0
          + \( V({\bf r}) - \bar{V} \)  \sigma_0
        \] \bra{\bf r},
 \label{eqn:H_W}
\end{align}
where ${\bf r}$ is the position of lattice sites and ${\bf e}_\mu$ 
($\mu = x,y,z$)
is the lattice vector in the $\mu$ direction.
 $\sigma_\mu$ are Pauli matrices, and $\sigma_0$ is the identity matrix.
 We use the hopping parameters $t = m_2 = 1$ as the energy unit.
 The on-site random potential $V({\bf r})$ uniformly distributes in $[-{W\over 2},{W\over 2}]$,
and the averaged potential $\bar{V}$ is subtracted so as to keep the energy of the Weyl point ($E=0$) unchanged.
 $m_0$ is the mass parameter controlling the topological property.
 We set the mass parameter $m_0$ so that a single pair of Weyl nodes appears roughly at $k_z = \pm{\frac{\pi}{2}}$, e.g., $m_0 = -1$ for $W = 0$ and $m_0 = -0.5$ for $W = 5$.
 The length unit is set to the lattice constant.

 We study the Hall and longitudinal transport in WSMs with 6-terminal Hall bar geometry (Fig.~\ref{fig:geometry}).
 The conductances are defined as
\begin{align}
 G_{xx} =  \frac{V_{xx}}{V_{xx}^2 + V\_{H}^2},\quad
 G\_{H} =  \frac{V\_{H}}{V_{xx}^2 + V\_{H}^2},
 \label{eqn:Gdef}
\end{align}
where the voltages $V_{xx}$ and $V\_{H}$ are numerically calculated using the recursive Green's function method \cite{Datta05quantum,Takane16disorder}.
 The pairs of choices of the electrodes (i.e., transport between two electrodes on the top or bottom and left or right) are equivalent and averaged in the following calculations.
 The electrodes are composed of ideal one-dimensional metallic wires, and we set the hopping between the sample and electrode sites $t^\prime = t$.
 Since we are interested in a thick limit of WSM thin films,
we impose the periodic boundary conditions in the $z$ direction for simplicity.

\iftitle
\section{Intrinsic anomalous Hall effect
} \label{sec:intrinsic}
\else
 \textit{Intrinsic anomalous Hall effect.}
\fi
\iftitle
\subsection{At the Weyl point} \label{sec:QuantizedRegime}
\fi
 First, we focus on the Weyl point $E=0$ of disordered WSM thin films.
 The clean WSM thin films are equivalent to the anomalous Hall insulators and 
are characterized by the Chern number \cite{Yoshimura16comparative}.
 For short-range disorder, the WSM encounters a semimetal-metal transition at a finite disorder strength $W\_c$~($\simeq 6$) \cite{Kobayashi14density, Kobayashi20ballistic}.
 Thus the WSM state,
i.e., vanishingly small density of states at $E=0$,
can survive even under disorder.

\begin{figure}[tbp]
 \centering
  \includegraphics[width=1\linewidth]{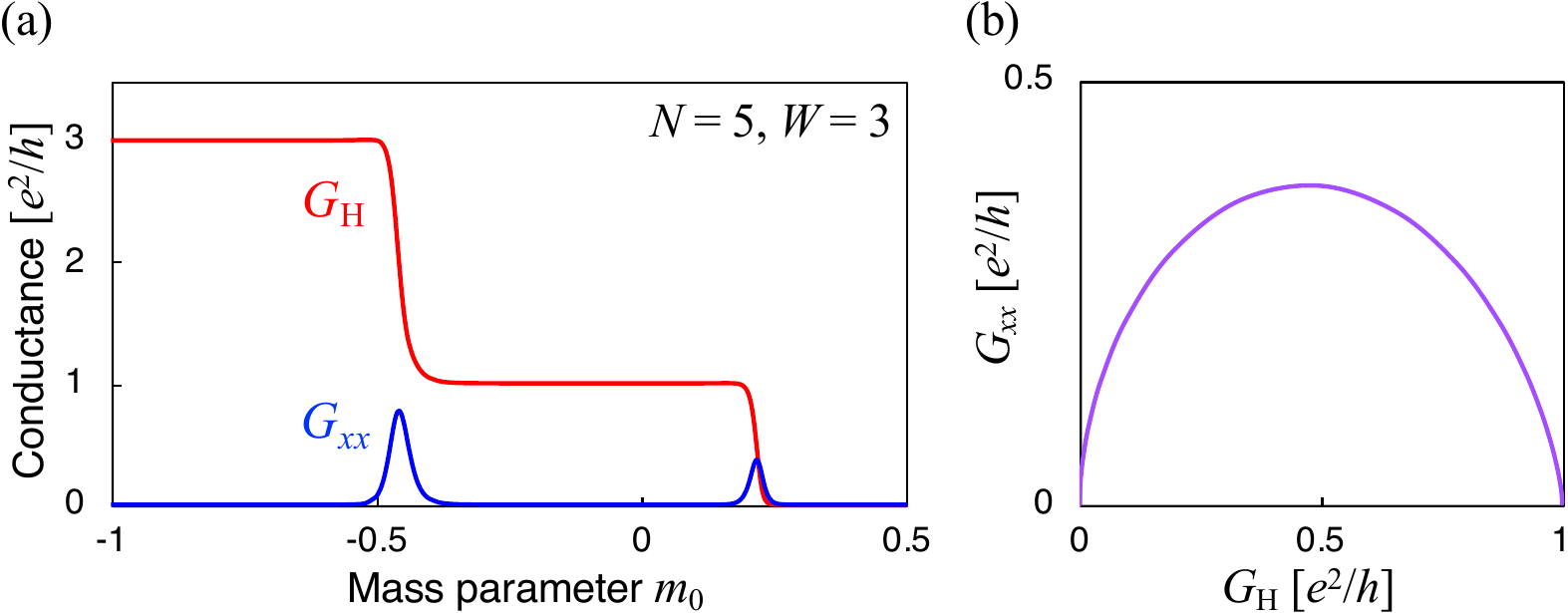}
 \vspace{-5mm}
\caption{\ifjpsj(Color online) \fi
  (a) Hall conductance $G\_H$ (red) and longitudinal conductance $G_{xx}$ (blue) as a function of mass $m_0$ at $E=0$.
  The size $L=60$, thickness $N=5$, and disorder strength $W=3$.
  (b) Relation between the conductances for the lowest plateau transition.
  We took the average over a sufficiently large number of samples,
  and the error bars are smaller than the line width.
}
\label{fig:QHE}
\end{figure}

 As long as the density of states at $E=0$ is small enough,
WSM thin films show the quantized Hall conductance as shown in Fig.~\ref{fig:QHE}(a).
 While the plateau transition points are governed by the effective mass $\tilde{m_0}(m_0,W)$ \cite{Shindou09effects,Goswami11quantum} instead of the bare $m_0$ under disorder,
the relation of the conductances, $G\_H$ and $G_{xx}$ [Fig.~\ref{fig:QHE}(b)], is insensitive to disorder and
seems the same as in QHIs \cite{Aoki87quantized,Huckestein95scaling}.
 We note that similar plateau transitions are also obtained as a function of disorder strength $W$ \cite{Takane16disorder}
since the effective mass $\tilde{m_0}$ is a function of $m_0$ and $W$.
 Therefore, for finite thickness, the intrinsic anomalous Hall effect in disordered WSM films is basically the same as the quantum anomalous Hall effect.
 In the thick limit or bulk WSMs, the discrete plateau transition turns into the continuous change of Hall conductance proportional to the separation of the Weyl nodes.

\begin{figure}[tbp]
 \centering
  \includegraphics[width=0.98\linewidth]{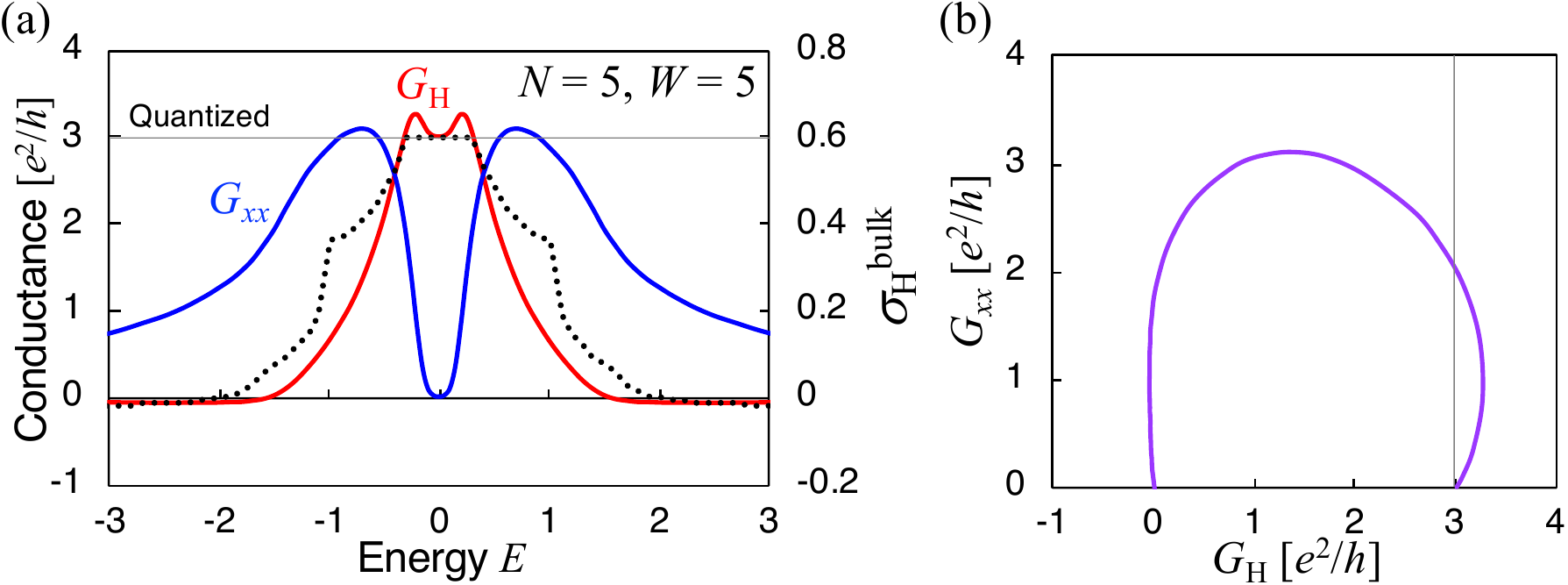}
 \vspace{-1mm}
\caption{\ifjpsj(Color online) \fi
  (a) Hall conductance $G\_H$ (red) and longitudinal conductance $G_{xx}$ (blue) as a function of Fermi energy $E$.
  The mass $m_0=-0.5$, size $L=60$, thickness $N=5$, and the disorder strength $W=5$.
  The dotted line is the Hall conductivity calculated from the Berry curvature in the clean limit. 
  (b) Relation between the conductances in (a).
  The error bars are smaller than the line width.
}
\label{fig:N5}
\end{figure}

\iftitle
\subsection{Energy dependence} \label{sec:EnergyDependence}
\fi
 The feature of the WSM distinctive from the QHI is the Fermi energy $E$ dependence of the density of states:
quadratic in 3D WSMs and vanishing only at $E=0$, whereas gapped in 2D QHIs.
 Here, we focus on the dependence of the conductances on Fermi energy.
 Sometimes the state at $E\ne 0$ is called Weyl metal,
but here we call the states within the energy range of the Weyl cones 
($|E|\lesssim t$ in the clean limit and $|E|\lesssim 0.6t$ for $W=5$)\cite{Kobayashi20ballistic}
as WSM, for simplicity.
 Under sufficiently strong disorder, where ordinary metallic states becomes diffusive,
the WSM states dominate the transport because they are insensitive to disorder especially near the Weyl point.
 Thus we consider a disorder strength $W=5$ slightly below the semimetal-metal transition $W\_c \simeq 6$
to focus on the transport of WSM states.
 The calculated Hall and longitudinal conductances as a function of Fermi energy are shown in Fig.~\ref{fig:N5}(a).
 At the Weyl point $E=0$,
the Hall conductance $G\_H$ has the quantized value $3$, and the longitudinal conductance $G_{xx}$ is almost zero.
 Here the quantized value of $G\_H$ is determined by how many 
wavenumbers $k_n=\frac{2\pi n}{N}$ cut the Fermi arc between $k_z =\pm \frac{\pi}{2}$:
$3$ for $N=5$ and $\frac{N}{2}$ in the large $N$ limit.
 For the Weyl metal $0<|E|\lesssim 0.6$, 
the Hall conductance $G\_H$ shows the double-peak structure
with a larger maximum value ($G\_H \simeq 3.3$ at $E=\pm 0.2$)
than the quantized value $3$, 
while the longitudinal conductance $G_{xx}$ increases with $|E|$.
 The Hall conductance decreases outside the WSM regime $|E| \gtrsim 0.6$ as the Fermi energy increases,
then vanishes for sufficiently large energy, i.e., for the diffusive metallic regime $|E| \gtrsim 1.5$.

 By plotting the relation between these conductances, Fig.~\ref{fig:N5}(b),
it becomes clear that the Hall conductance increases from the quantized value as the longitudinal conductance increases from zero.
 This curve corresponds to an energy-driven crossover from the Weyl point $E=0$ ($G\_H=3$ and $G_{xx}=0$)
to the Hall conductance peak in the WSM regime ($G\_H > 3$),
then to the diffusive metal or the Anderson insulator ($G\_H\simeq 0$ and $G_{xx}>0$).
 The point $G\_H=G_{xx}=0$ corresponds to the outside of the energy band $|E|\gtrsim 6$.
 Reflecting the increase of the Hall conductance,
the curve has a positive slope around the Weyl point $(G\_H,G_{xx})=(3,0)$,
in contrast to that in the QHIs, which has a negative slope around $(G\_H,G_{xx})=(1,0)$ [see Fig.~\ref{fig:QHE}(b)].
 This relation means that the deviation of the Hall conductance from the quantized value behaves as 
$\Delta G\_H(E) \equiv G\_H(E) - G\_H(0) \sim G_{xx}(E)$ 
in the vicinity of the Weyl point, 
and thus the Hall angle $G\_H/G_{xx}$ keeps a large value
even if the Fermi energy deviates from the Weyl point and if the system is disordered.

\begin{figure}[tbp]
 \centering
  \includegraphics[width=0.98\linewidth]{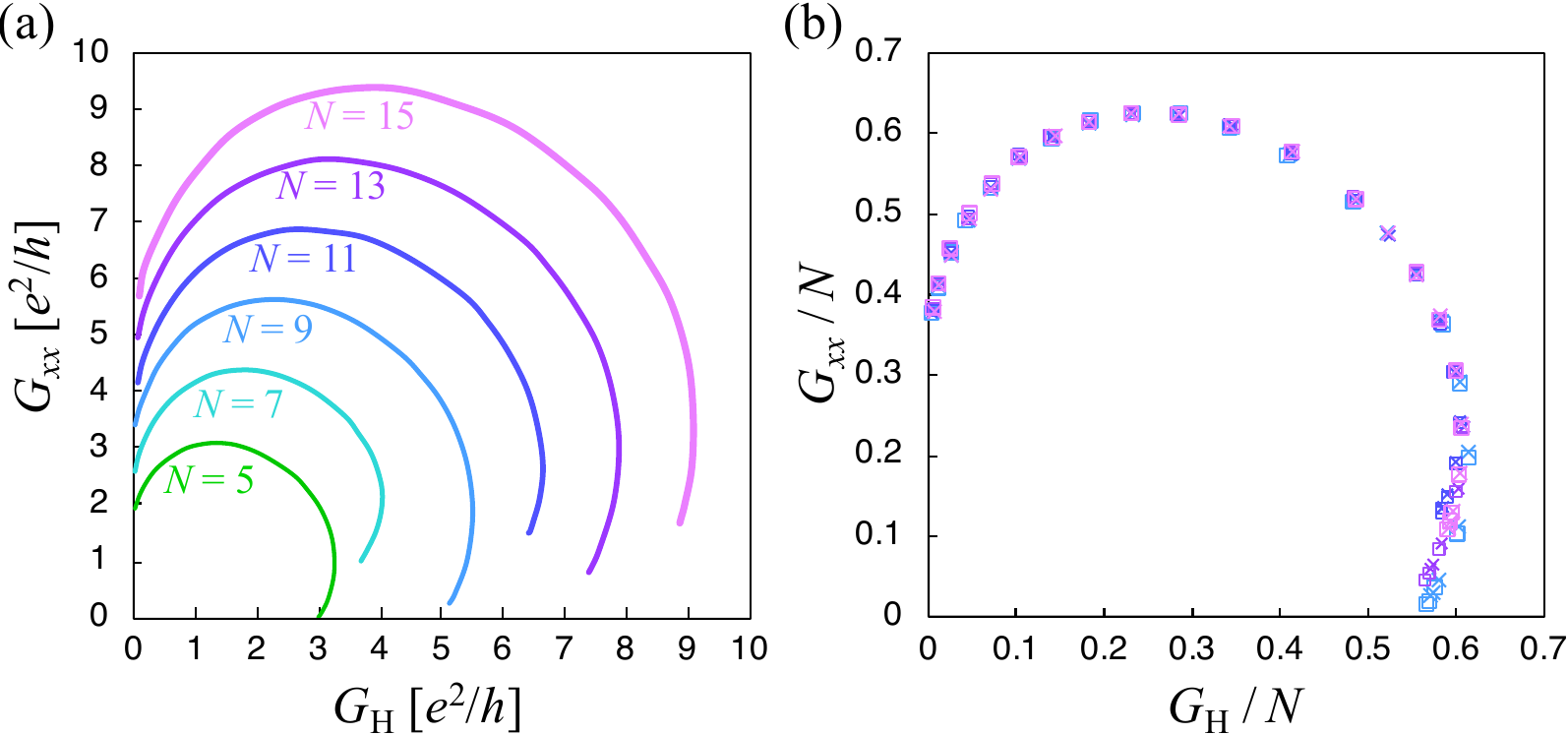}
 \vspace{-1mm}
\caption{\ifjpsj(Color online) \fi
  (a) Relation of the energy-dependent conductances in magnetic Weyl semimetal films with thicknesses $N=5,7,9,11,13$, and $15$.
  The area size $L=60$, disorder strength $W=5$ and $m_0=-0.5$.
  (b) Relation between the scaled conductances $G\_{H}/N$ and $G_{xx}/N$.
  The sizes are $L=60$ (cross) and $90$ (square) with thicknesses $N=9,11,13$, and $15$.
  The statistical error bars are smaller than the symbols.
}
\label{fig:relationScaled}
\end{figure}

 Next, we investigate the system-size dependence of the relation.
 Figure~\ref{fig:relationScaled}(a) shows the relations for different thicknesses.
 The conductances are almost proportional to the thickness $N$, and the relation keeps its form when the thickness increases.
 On the other hand, they are almost independent of the area size of the system $L$,
at least up to $L=90 \approx 0.1\mathrm{\mu m}$.
 Therefore, by scaling the conductances with $N$ (which has the same dimension as conductivity), we obtain an almost single curve for a certain strength of disorder Fig.~\ref{fig:relationScaled}(b).
 In the thick WSM limit $N\to \infty$, 
we expect the curve starts from 
$(G\_H/N,G_{xx}/N) \to (0.5,0)$.
 On the other hand, the data for the maxima of the Hall conductance are almost converged around $(G\_H/N,G_{xx}/N)=(0.6,0.25)$.
 Thus we expect the increasing behavior survives in a thick WSM.


\begin{figure}[tbp]
 \centering
  \includegraphics[width=0.98\linewidth]{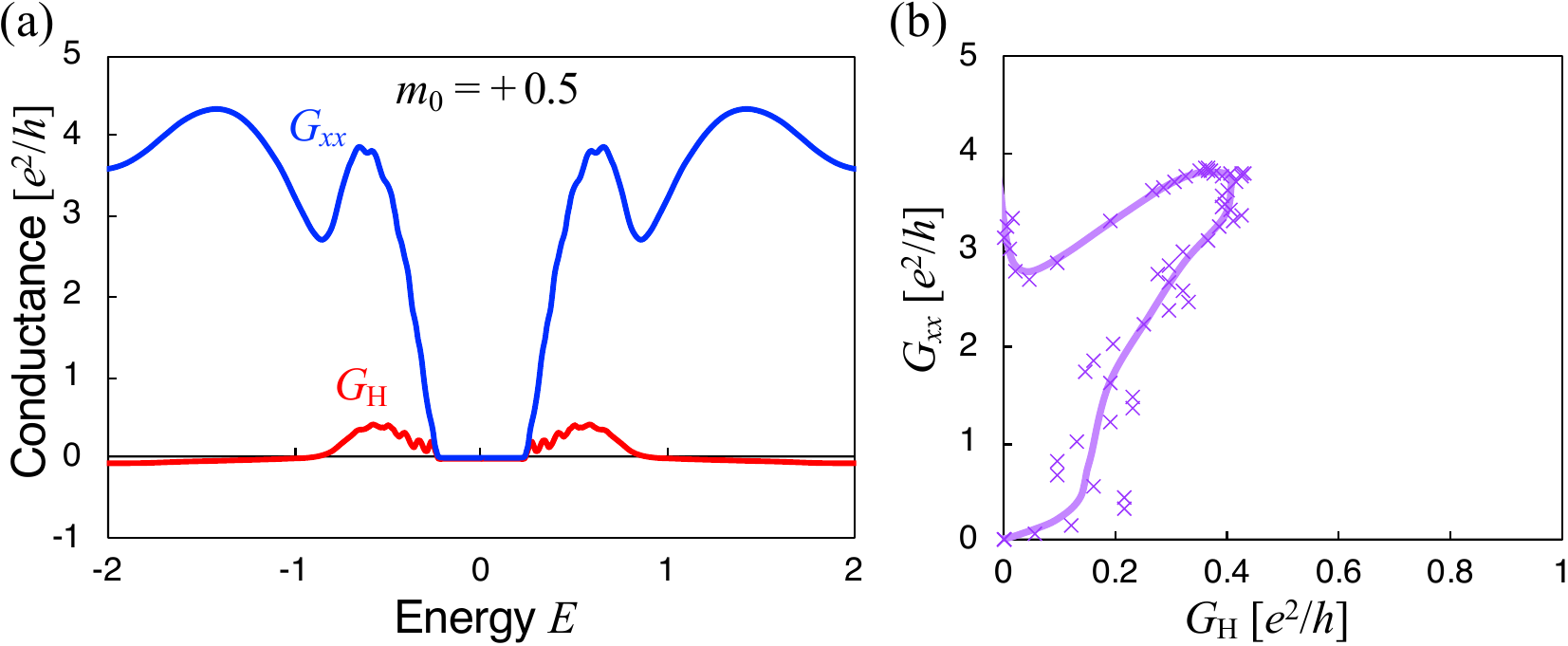}
 \vspace{-3mm}
\caption{\ifjpsj(Color online) \fi
  (a) Hall conductance $G\_H$ (red) and longitudinal conductance $G_{xx}$ (blue) as a function of Fermi energy $E$.
  The mass $m_0=+0.5$, size $L=30$, thickness $N=5$, and disorder strength $W=3$.
  The Hall conductivity calculated from the Berry curvature in the clean limit is negligibly small. 
  This Hall conductance peak is rather fragile than that in the gapless WSM;
  it tends to vanish for a large gap or strong disorder.
  (b) Relation between the conductances in (a).
  The ensemble averages over 150,000 samples are taken for each data point, and 
  the error bars are smaller than the symbols.
  The solid line is a guide to the eyes (moving average).
}
\label{fig:gapped}
\end{figure}

 The increasing behavior of the Hall conductance $G\_H$ with increasing longitudinal conductance $G_{xx}$ cannot be explained by the Berry curvature or linear response theory in the clean bulk [dotted line in Fig.~\ref{fig:N5}(a)].
 For a simple WSM, it was predicted that a finite Hall conductance arises even if the Weyl cones are gapped \cite{Burkov15chiral}.
 In fact, we have confirmed that the Hall conductance can be finite for gapped Weyl states in our model.
 For a small positive effective mass $\tilde{m}_0 > 0$, 
the bulk Weyl cones are gapped,
and the chiral edge states in thin films are absent.
 Figure~\ref{fig:gapped}(a) shows the Hall and longitudinal conductances for the gapped state $m_0 = +0.5$ with $W=3$.
 Although the system is topologically trivial,
the Hall conductance becomes finite and shows the double-peak structure around the gap ($G\_H \simeq 0.4$ at $E=\pm 0.5$).
 By plotting the relation between $G\_H$ and $G_{xx}$ [Fig.~\ref{fig:gapped}(b)],
one can see that the Hall conductance increases with increasing longitudinal conductance.
 Since the relation is similar to that for the additional Hall conductance in gapless WSMs
[the relation around $(G\_H, G_{xx})$ = $(3,0)$ in Fig.~\ref{fig:N5}(b)],
double-peak structures around $E = 0$ for gapless [Fig.~\ref{fig:N5}(a)] and gapped [Fig.~\ref{fig:gapped}(a)] WSMs are expected to have a common origin.
 Thus, the double-peak structures
can be interpreted as the contributions to Hall conductance from the gapped subbands of Weyl cones, 
which arise in gapless WSMs for $k_z$'s around but not exactly at the Weyl nodes $k_z = \pm k_0$.

 Here we comment on the disorder strength.
 The presence of disorder will be essential for the additional Hall conductance peaks around the Weyl point,
since the bulk Berry curvature in the absence of disorder does not show the peak structure.
 However, the obtained additional peaks
become prominent in the weak disorder limit and 
are suppressed as increasing the disorder strength $W$.
 Although this behavior is not intuitive,
similar behavior is reported for the valley Hall conductivity in graphene \cite{Ando15theory}, calculated by the self-consistent Born approximation.
 Thus the enhancement of the Hall conductance in the weak disorder limit
is not a numerical artifact
but can be a generic feature of Dirac electron systems.
 Meanwhile,
in the clean limit or weak disorder such that the mean free path exceeds the system size,
the conducting states are ballistic,
and the conductances are proportional to the number of current-carrying states ($G\_H,G_{xx}\sim L$).
 Therefore, the scale-insensitive relation found
in Fig.~\ref{fig:relationScaled}(b) holds only for a sufficiently disordered (i.e., diffusive) regime.

\iftitle
 \section{Extrinsic anomalous Hall effect}
 \label{sec:extrinsic}
\else
 \textit{Extrinsic anomalous Hall effect.}
\fi
 The extrinsic (i.e., impurity-induced) anomalous Hall conductance is typically proportional to the longitudinal conductance and should be vanishing at the Weyl point $E=0$, where $G_{xx}=0$.
 Furthermore, even for Weyl metals $|E|>0$, the extrinsic contribution is predicted \cite{Burkov14anomalous,footnote} to be buried in the intrinsic contribution.
\iftitle
 In this section, 
\else
 Here,
\fi
 we carefully extract the extrinsic contribution to the anomalous Hall effect in ferromagnetic WSMs
and show the relation between the extrinsic contribution and the longitudinal conductance.

\iftitle
\subsection{Dirac semimetal model} \label{sec:extrinsicModel}
\fi
 Since the extrinsic anomalous Hall effect is maximized in half-metallic systems,
we employ a model for half-metallic magnetic WSMs, 
which is realized by introducing a strong exchange interaction into topological Dirac semimetals.
 The Hamiltonian for a topological Dirac semimetal is typically written as \cite{Wang12dirac}
\begin{align}
 H\_{D} &=  \sum_{\bf r}
         \[ {it \over 2} 
           \(
              \ket{{\bf r}+{\bf e}_x}
               \tau_x s_z
              \bra{\bf r}
            + \ket{{\bf r}+{\bf e}_y}
               \tau_y s_0
              \bra{\bf r}
           \)
           + \textrm{H.c.}
         \] \nonumber \\
   &+  \sum_{\bf r}
        \sum_{\mu=x,y,z}
         \[ \ket{{\bf r}+{\bf e}_\mu}
           \(
              -{m_2 \over 2} \tau_z s_0
           \)
           \bra{\bf r}  + \textrm{H.c.}
         \]   \nonumber \\
   & + \sum_{\bf r} \ket{\bf r}
        \[m_0 \tau_z s_0
          + U({\bf r}) \tau_0 s_0 
          - JM_z \tau_0 s_z
        \] \bra{\bf r},
 \label{eqn:H_TDS}
\end{align}
where $\tau_i$ and $s_i$ with $i=x,y,z$ are Pauli matrices corresponding to orbital and spin degrees of freedom, respectively, and with $i=0$ are identity matrices.
 We set $t=2$ and $m_2=1$.
 We consider the short-ranged impurity potential $U({\bf r})$ with potential height $W\_{imp}$,
which is randomly distributed on the lattice sites with the density $\rho\_{imp}$.
 The exchange interaction $J$ and magnetization $M_z$ are set so that the spin-up and -down bands split sufficiently ($JM_z = 8$).
 We focus on the spin-up bands, and 
the energy is measured from the Weyl point.

 The spin-dependent scattering of strength $W_s$ is introduced in the form of $H_s \propto W_s \({\bf S} \times {\bf p}\)\cdot \nabla U$ \cite{Nikolic07extrinsically,Kobayashi21ferromagnetic}, as
\begin{align}
 H_s &=  \sum_{\bf r} \ket{\bf r}
          \sum_{\mu,\nu,\gamma=x,y,z} \sum_{m,n=\pm 1}
          (- i W_s) \varepsilon_{\mu\nu\gamma} m n S_\gamma \nonumber \\
     &  \times \[   U({\bf r} + m{\bf e}_\mu) - U({\bf r} + n{\bf e}_\nu) 
          \] \bra{{\bf r} + m{\bf e}_\mu + n{\bf e}_\nu},
 \label{eqn:H_s}
\end{align}
where the spin operators $S_{x,z}=\tau_0 s_{x,z}$ and $S_{y}=\tau_z s_{y}$.

\iftitle
 \subsection{Extrinsic contributions} \label{sec:extrinsicResult}
\fi

 We investigate the Hall conductance in the ferromagnetic WSMs with impurities.
 To focus on the extrinsic anomalous Hall effect,
we consider sufficiently weak disorder, where the localization length is larger than the system size
and $G_{xx}$ (and thus the extrinsic contribution to $G\_H$) becomes large.
 Here we set the potential height $W\_{imp}=4$ and density $\rho\_{imp}=5\%$,
which are efficient to obtain large extrinsic anomalous Hall effect.

\begin{figure}[tbp]
 \centering
  \includegraphics[width=0.98\linewidth]{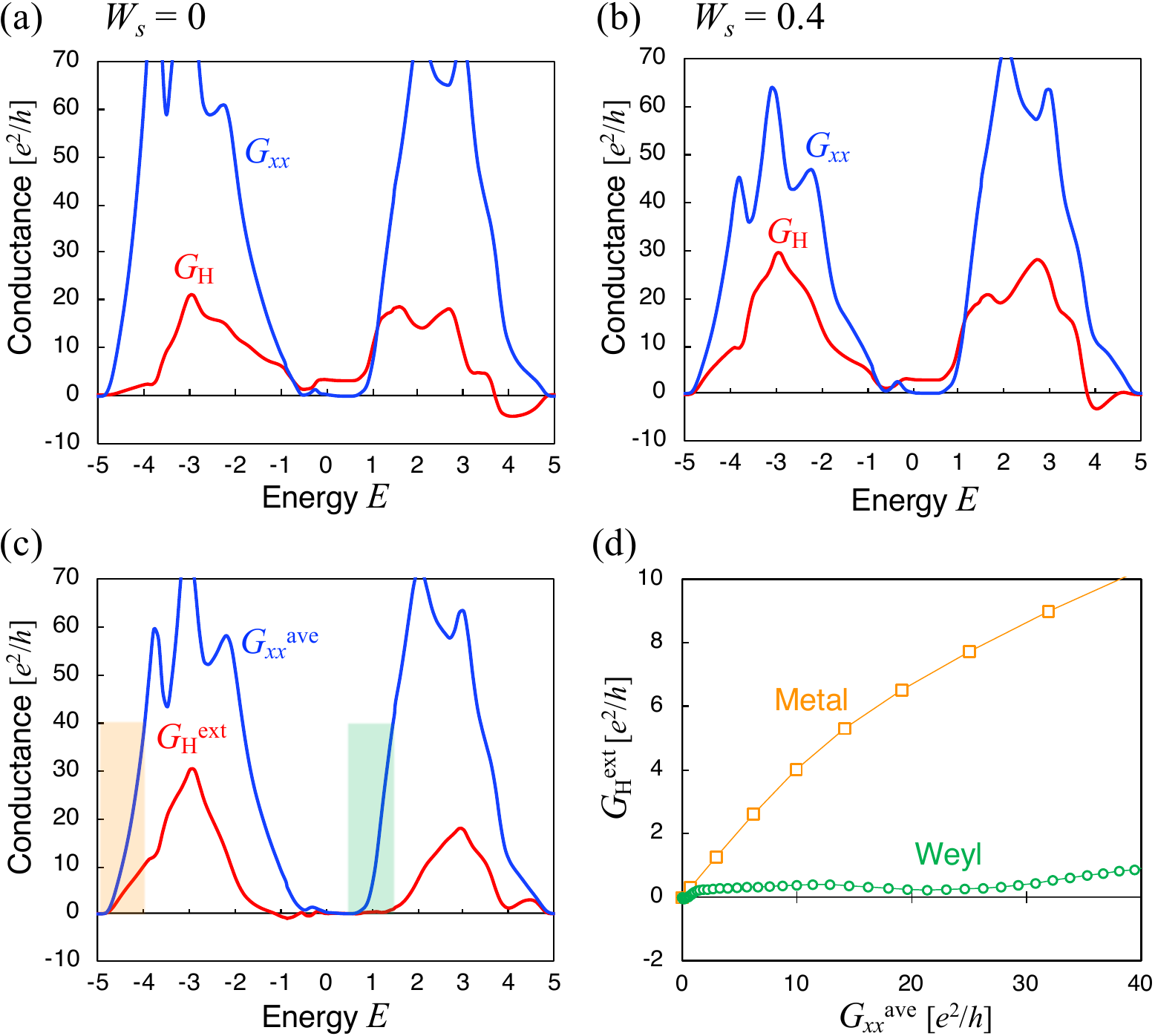}
 \vspace{-2mm}
\caption{\ifjpsj(Color online) \fi
  (a)(b) Hall conductance $G\_H$ (red) and longitudinal conductance $G_{xx}$ (blue) as functions of Fermi energy $E$ under impurities of height $W\_{imp}=4$ with density $\rho\_{imp}=5\%$.
  The strength of the spin scattering is $W_s = 0$ (a) and $W_s = 0.4$ (b).
  The system size $L = 30$, thickness $N = 5$, and the mass $m_0=-0.5$.
  (c) Longitudinal conductance $G_{xx}^{\mathrm{ave}}(E) = [G_{xx}(E,0.4) + G_{xx}(E,-0.4)]/2$ (blue) and extrinsic contribution of Hall conductance $G\_H^{\mathrm{ext}}$ (red) calculated from $W_s=0$ (a), $0.4$ (b), and $-0.4$ (not shown).
  (d) Relations of conductances for the Weyl semimetal state (green circle) corresponding to the green shadowed region in (c) and for the ordinary metallic state (orange square) corresponding to the orange shadowed region.
  The statistical error bars are smaller than the symbols.
}
\label{fig:extrinsic}
\end{figure}

 In the absence of the spin scattering $W_s=0$,
we see the intrinsic contributions to the anomalous Hall effect as shown in Fig.~\ref{fig:extrinsic}(a).
 Since the disorder strength is weak compared with the white-noise disorder of $W=5$ discussed in Figs.~\ref{fig:N5} and~\ref{fig:relationScaled},
an energy gap ($|E|\lesssim 0.6$) and a quantized plateau of the Hall conductance remain for a thin film ($N=5$).
 The small dip and peak of $G\_H$ and $G_{xx}$, respectively, around $E=-0.5$ are coming from the impurity level,
which we do not go into the details here.
 The large Hall conductance around $|E|\simeq 3$, which locates in the metallic regime, is due to the large longitudinal conductance \cite{Nagaosa10anomalous} and strong exchange coupling.

 By introducing a finite spin scattering $W_s=0.4$,
both the Hall and longitudinal conductances change as shown in Fig.~\ref{fig:extrinsic}(b).
 Since the obtained Hall conductance is a sum of intrinsic and extrinsic contributions,
we extract the extrinsic contribution by defining
\begin{align}
 G\_H^{\mathrm{ext}}(E,W_{\!s}) &= G\_H(E,W_{\!s}) \nonumber\\
    & - G\_H(E,0) \frac{G_{xx}(E,W_{\!s})\!+\!|G\_H(E,W_{\!s})|}{G_{xx}(E,0)\!+\!|G\_H(E,0)|},
 \label{eqn:GHext}
\end{align}
where the second term is an estimation of the intrinsic contribution
($|G\_H|$'s are added to avoid the divergence when $G_{xx}=0$ and not important for $G_{xx}>0$, which we are interested in).
 For the sake of numerical accuracy, we have calculated the average $G\_H^{\mathrm{ext}}(E) = [G\_H^{\mathrm{ext}}(E,0.4) - G\_H^{\mathrm{ext}}(E,-0.4)]/2$, where $G\_H^{\mathrm{ext}}(E,-0.4)$ gives a similar value to the opposite sign of $G\_H^{\mathrm{ext}}(E,0.4)$.
 The extracted extrinsic contribution $G\_H^{\mathrm{ext}}$ is plotted in Fig.~\ref{fig:extrinsic}(c).
 Compared to the extrinsic Hall conductance $G\_H^{\mathrm{ext}}$ in the metallic regime 
($-5 \leq E \leq -4$, orange shadowed region),
that for the WSM regime 
($0.5 \leq E \leq 1.5$, green shadowed region)
looks suppressed.

 The suppression of the extrinsic effect in WSMs becomes clear by plotting the relation of conductances.
 As shown in Fig.~\ref{fig:extrinsic}(d),
the extrinsic Hall conductance in the WSM state is slightly increasing with increasing longitudinal conductance, but its Hall angle 
$G\_H^{\mathrm{ext}}/G_{xx}^{\mathrm{ave}}$
is significantly smaller than that in the metallic state.
 This small Hall angle implies
that the impurity-induced deviation of the Hall conductance will be 
smaller than the additional contribution from the subbands of Weyl cones.
 Thus the extrinsic contribution,
which can be either additive or subtractive to the intrinsic contribution,
does not change the increasing behavior in Fig.~\ref{fig:relationScaled}(b) qualitatively
even with a large longitudinal conductance, i.e., a larger system size.

\iftitle
\section{Conclusions} \label{sec:conclusion}
\else
 \textit{Conclusions.}
\fi
 We have studied the relation between the Hall conductance and longitudinal conductance in disordered WSMs.
 We have first shown that the WSM thin films at $E=0$, where the density of states vanishes, reproduces the same relation as for the QHIs even under strong disorder.
 In contrast, in WSMs with a finite density of states, i.e., when the Fermi energy shifts from the Weyl point, a specific relation for the WSM is found:
the Hall conductance $G\_H$ increases from the semi-quantized value with increasing longitudinal conductance $G_{xx}$.
 The increase of the Hall conductance is considered to be coming from the subbands of the Weyl cones
and is not predictable only from the bulk Berry curvature in the clean limit.
 We have investigated the system-size dependence of the conductances and 
found that 
the conductances linearly scale with the thickness $N$, and 
their specific relation 
 (i.e., the increasing behavior of $G\_H/N$ with $G_{xx}/N$)
survives for thick WSMs.
 We have also investigated the extrinsic (skew scattering induced) anomalous Hall effect in ferromagnetic WSMs.
 By carefully extracting the extrinsic contributions,
we have confirmed that the extrinsic anomalous Hall effect is nearly independent of the longitudinal conductance and significantly smaller than the intrinsic contributions in the WSM state.
 These results imply the robustness of the Hall angle against disorder and deviation of the energy from the Weyl point; they will explain the large Hall angle in doped WSMs or disordered Weyl metals.

\begin{acknowledgments}
 We thank M. Koshino for valuable discussions.
 This work was supported by 
the Japan Society for the Promotion of Science KAKENHI (Grant Nos.~%
JP19K14607 
and
JP20H01830) 
and
by CREST, Japan Science and Technology Agency (Grant No.~JPMJCR18T2).
\end{acknowledgments}


\bibliography{GHGxx}

\begin{thebibliography}{29}
\bibitem [{Note1()}]{footnote}%
  \BibitemOpen
  \bibinfo {note} {The skew scattering was not taken into account in
  Ref.~\onlinecite{Burkov14anomalous}}\BibitemShut {NoStop}%
\end{thebibliography}

\end{document}